\documentclass[useAMS,usenatbib]{mnras}
\usepackage{times}
\usepackage{url}
\usepackage{graphicx}
\usepackage[usenames]{color}
\DeclareGraphicsExtensions{.pdf,.png,.jpg,.mps,.eps,.ps}
\usepackage{amsmath}
\usepackage{natbib}
\def\unit #1{\,{\rm #1}}

\newcommand\cmsqi{\rm \,\unit{cm^{-2}}}

\newcommand\kev{\rm \,\unit{keV}}

\newcommand\ergs{\rm \,\unit{erg\,s^{-1}}}

\newcommand\lunit{\rm \,erg \,s^{-1}}

\newcommand\xiunit{\rm \,erg\,cm\,s^{-1}}

\newcommand\nh{ N_{\rm H}}

\newcommand\ks{\, \rm ks}

\newcommand\dc{\, \Delta\chi^2}

\newcommand\cd{\,\rm \chi^2/dof}

\newcommand\ev{\unit{\, eV}}

\newcommand\fe{\rm Fe\, K\alpha}

\newcommand\lbol{L_{\rm \, bol}}

\newcommand\msol{M_{\odot}}

\newcommand\chandra{{\it Chandra}}

\newcommand\xmm{{\it XMM-Newton}}

\newcommand\suzaku{{\it Suzaku}}
\newcommand\nustar{{\it Nustar}}

\usepackage{graphicx}
\title{Investigating the origin of the Fe emission lines of the Seyfert 1 galaxy Mrk~205.}
\author[Laha et al.]{Sibasish Laha$^{1}$\thanks{sib.laha@gmail.com, slaha@ucsd.edu}, Ritesh Ghosh$^2$, Shruti Tripathi$^3$ and Matteo Guainazzi$^{4}$ \\
$^{1}$ University of California, San Diego, Center for Astrophysics and Space Sciences, 9500 Gilman Dr, La Jolla, CA 92093-0424, USA.\\
$^2$Visva-Bharati University, Santiniketan, Bolpur 731235, West Bengal, India. \\
$^3$ Department of Astronomy \& Physics, Saint Mary's University, Halifax, NS, B3H 3C3, Canada\\
$^{4}$European Space Research and Technology Centre, Keplerlaan 1, 2201 AZ Noordwijk, Netherlands.\\
}

\date{\today}
\begin{document}
\pagerange{\pageref{firstpage}--\pageref{lastpage}} \pubyear{2013}

% \def\LaTeX{L\kern-.36em\raise.3ex\hbox{a}\kern-.15em
% T\kern-.1667em\lower.7ex\hbox{E}\kern-.125emX}

% \newtheorem{theorem}{Theorem}[section]

% \label{firstpage}

\maketitle
\label{firstpage}
\begin{abstract}

  We have investigated the nature and origin of the Fe K emission lines in Mrk 205 using observations with \suzaku{} and \xmm{}, aiming to resolve the ambiguity between a broad emission line and multiple unresolved lines of higher ionization. We detect the presence of a narrow Fe K$\alpha$ emission line along with a broad band Compton reflection hump at energies $E>10\kev$. These are consistent with reflected emission of hard X-ray photons off a Compton thick material of $\nh\ge 2.15\times 10^{24}\cmsqi$. In addition we detect a partially covering ionized absorption with ionization parameter $\log(\xi/\xiunit)=1.9_{-0.5}^{+0.1}$, column density $\nh=(5.6_{-1.9}^{+2.0})\times 10^{22}\cmsqi$ and a covering factor of $0.22_{-0.06}^{+0.09}$. We detect the presence of emission arising out of ionized disk reflection contributing in the soft and the hard X-rays consistently in all the observations. We however, could not definitely ascertain the presence of a relativistically broadened Fe line in the X-ray spectra. Using relativistic reflection models, we found that the data are unable to statistically distinguish between the scenarios when the super-massive black hole is non-rotating and when it is maximally spinning. Using the disk reflection model we also find that the accretion disk of the AGN may be truncated at a distance $6R_{\rm G}<R<12R_{\rm G}$, which may suggest why there may not be any broad Fe line. The Eddington rate of the source is low ($\lambda_{\rm Edd}=0.03$), which points to an inefficient accretion, possibly due to a truncated disk.

 \end{abstract}

\begin{keywords}
	galaxies: individual Mrk~205--galaxies: Seyfert--galaxies: active--galaxies: kinematics and dynamics. 

\end{keywords}

\vspace{0.5cm}

\section{INTRODUCTION}

The most well accepted scenario of X-ray spectral emission in active galactic nuclei (AGN) is that of an optically-thin and hot ($T\sim 10^9$ K) corona \citep{1993ApJ...413..507H,1994ApJ...432L..95H} upscattering the seed UV photons emitted from the accretion disk \citep{1973A&A....24..337S} around the central super-massive black hole (SMBH), resulting in a power law spectrum. It is also believed that the X-ray photons from the corona are reflected off the cold nuclear matter (popularly called torus) as well as the ionized accretion disk to produce fluorescent emission lines of Fe of various ionic stages. The relatively higher abundance of Fe coupled with a higher fluorescent yield, makes the $\fe$ emission line the most prominent and ubiquitous of all emission lines in the X-ray spectra of AGN.

The $\fe$ line was detected for the first time by \citet{1978ApJ...220..790M} using {\it OSO-8} observations of Centaurus-A. With the advent of the latest generation X-ray telescopes such as \xmm{}, \suzaku{} and \chandra{}, the presence of the $\fe$ line in AGN X-ray spectra has been found to be almost ubiquitous \citep{2011ApJ...727...19F}. The neutral $\fe$ emission line with a rest frame energy of $6.4\kev$, arising out of the distant reflection off neutral medium (torus) is mostly narrow ($\sigma\le 100\ev$) and is associated with a Compton reflection hump peaking at energies $>10\kev$. On the other hand when the X-ray coronal photons are reflected off the ionized accretion disk, the rest frame energy of the $\fe$ emission line may shift to higher energies ($6.40<E<6.96\kev$). {In addition to that, if the reflection happens from the inner regions of the accretion disk where the matter is in relativistic motion around the SMBH, the emitted $\fe$ line profile becomes broad and skewed with a red wing extending towards lower X-ray energies due to gravitational redshift and transverse Doppler redshift \citep[see e.g.,][]{1989MNRAS.238..729F}. These relativistic $\fe$ emission lines sometimes originate within a few gravitational radii from the central object and hence serve as important probes of the physical processes taking place in the inner most regions of the AGN \citep{2010A&A...524A..50D,2011MNRAS.414.1269W,2014MNRAS.439.2307F}.} Thus both the narrow and broad $\fe$ emission lines in the X-ray spectra give us important clues to the hitherto spatially unresolved inner regions of AGN in the vicinity of the SMBH. However, some authors have cautioned that warm absorbers and/or neutral absorbers with sufficiently high column density may modify the $3-7\kev$ energy range of the X-ray spectra of AGN, mimicking a broad relativistic $\fe$ line with a red wing \citep[see e.g.,][]{2004ApJ...602..648R,2008A&A...483..437M,2009A&ARv..17...47T}. Similarly, a blend of the helium and hydrogen-like iron at $6.7-6.9\kev$ can be misinterpretated as a broad diskline profile detected at high inclination, particularly in spectra at moderate (CCD) energy resolution \citep{2003ApJ...596...85Y,2007A&A...470...73L}.

\citet{2010A&A...524A..50D} studied the nature of the {relativistically broadened $\fe$ emission line in a sample of 31 Type-1 AGN}. The authors found that the occurence probability of a broad $\fe$ emission line is $\sim36\%$ with an average equivalent width of $\sim 100\ev$. The average accretion disk inclination detected in the AGN was $28 \pm 5^{\circ}$. A recent sample study on the narrow $\fe$ emission line and its link with molecular torus was carried out by \citet{2014A&A...567A.142R}. The authors detected the line ubiquitously in a sample of 24 AGN. More interestingly, with simultaneous mid-infra-red (MIR) studies they found that the narrow $\fe$ line is produced by the same material that is responsible for the MIR emission in AGN, implying that the molecular torus is an important reprocessor of the AGN primary emission.

Mrk~205 is a nearby ($z=0.071$) low luminosity radio quiet quasar with an intriguing $\fe$ emission line complex. \citet{2001A&A...365L.134R} studied this source with \xmm{} and detected a narrow $\fe$ as well as a {moderately broad (rms width $\sigma=300\ev$) $\fe$ emission lines at $6.4\kev$ and $6.7\kev$ respectively, in the rest frame}. The authors suggested that the broad line arises from X-ray reflection off the surface of a highly ionized accretion disk, while the narrow component is due to the reflection of X-ray photons off a distant neutral matter, possibly the torus. However, the authors could not affirmatively distinguish whether the {`broadened line'} did originate from a relativistic accretion disk, or it's a blend of higher ionization Fe emission lines. A similar study of Mrk~205 carried out by \citet{2003MNRAS.343.1241P} using \xmm{} observations found that the Fe K emission lines do not present any strong evidence for reprocessing in the inner, relativistic parts of accretion disks. {\citet{2012MNRAS.426.2522P} studied the \suzaku{} spectra of the source and could not detect any broad Fe emission line. Using the same \suzaku{} observation \citet{2016PASJ...68S..27I} detected a partially covering ionized absorber which mimics a relativistically broadened Fe line.} The detection and origin of the broad $\fe$ emission line of Mrk~205 is therefore still a matter of debate.

In recent years there has been a rapid development in our understanding of the broad $\fe$ line in terms of relativistic disk reflection models such as {\it relxill} \citep{2013ApJ...768..146G} which consistently models the broad Fe K feature along with the soft X-ray excess ($<2\kev$) and the hard X-ray bump ($>10\kev$) due to disk reflection. Moreover, more sophisticated models for reprocessing by optically thick cold matter have been developed, such as {\it MYTorus} \citep{2009MNRAS.397.1549M} which consistently models the narrow $\fe$ emission line with the Compton reflection hump at $E>10\kev$. Thus, spectral coverage in the $10-30\kev$ energy band is highly important to break the degeneracy between the broad $\fe$ lines and blend of higher ionization $\fe$ and other Fe lines. With the availability of the \suzaku{} broadband spectral data for the source Mrk~205, and with the recent physical models, we are now critically poised to investigate the origin of the iron complex in Mrk~205 in detail. In this paper we carry out a detailed X-ray spectral analysis of the source Mrk~205 using \suzaku{} and \xmm{} telescopes focussing on the origin of the narrow and the broad Fe K emission lines  using realistic broad band models, and shedding light on the reprocessing medium around the SMBH.

The paper is arranged as follows: Section \ref{sec:obs} describes the observation and data reprocessing. Section \ref{sec:analysis} lists the steps taken in the spectral analysis. Section \ref{sec:results}  discusses the results followed by conclusions in Section \ref{sec:conclusions}.    \\

%%%%%%%%%%%%%%%%%%%%%%%%%%%%%%%%%%%%%%%%%%%%%%%    Observation and data reduction

\section{Observation and data reprocessing}\label{sec:obs}

Mrk~205 was observed by \xmm{} on two occasions in 2000 and 2006 for $15\ks$ and $100\ks$ respectively, and once by \suzaku{} in 2010 for $101\ks$. See Table \ref{Table:obs} for details and the short notation of the observation ids. The \suzaku{} reprocessing and spectral extraction were carried out following the steps described in \citet{2018MNRAS.480.1522L} and \citet{2016MNRAS.456..554G}. There is no pile up in the XIS spectra of the source. We extracted the hard X-ray spectral data using the HXD- PINXBPI tool from the PIN cleaned events and the pseudo-event lists generated by the AEPIPELINE tool. See \citet{2016MNRAS.456..554G} for a description of the methods involved in extracting the PIN spectrum.

 The EPIC-pn data from \xmm{} were reduced using the scientific
analysis system (SAS) software (version 15) with the task {\it
  epchain} and using the latest calibration database. We used EPIC-pn data because
of their higher signal to noise ratio as compared to MOS in the Fe-K energy band, our region of interest. For filtering and spectral extraction we followed the methods described in \citet{2014MNRAS.441.2613L}. 

The \suzaku{} XIS and the {\it PIN} spectra were grouped by a minimum counts of 200 and 20 per energy bin, respectively, using the command {\it grppha} in the {\it HEASOFT} software. The \xmm{} spectra were grouped by a minimum of 20 counts per channel and a maximum of five resolution elements using the command {\it specgroup} in SAS. The {\it XSPEC} \citep{1996ASPC..101...17A} software was used for spectral analysis. All errors quoted on the fitted parameters reflect the $90\%$ confidence interval for one interesting parameter, corresponding to $\Delta \chi^2=2.7$ \citep{1976ApJ...208..177L}.

%%%%%%%%%%%%%%%%%%%%%%%%%%%%%%%%%%%%%%%%%%%%%%%%   Data analysis

\section{Data analysis}\label{sec:analysis}

\begin{figure*}
  \centering 

\hbox{
\includegraphics[width=5.5cm,angle=-90]{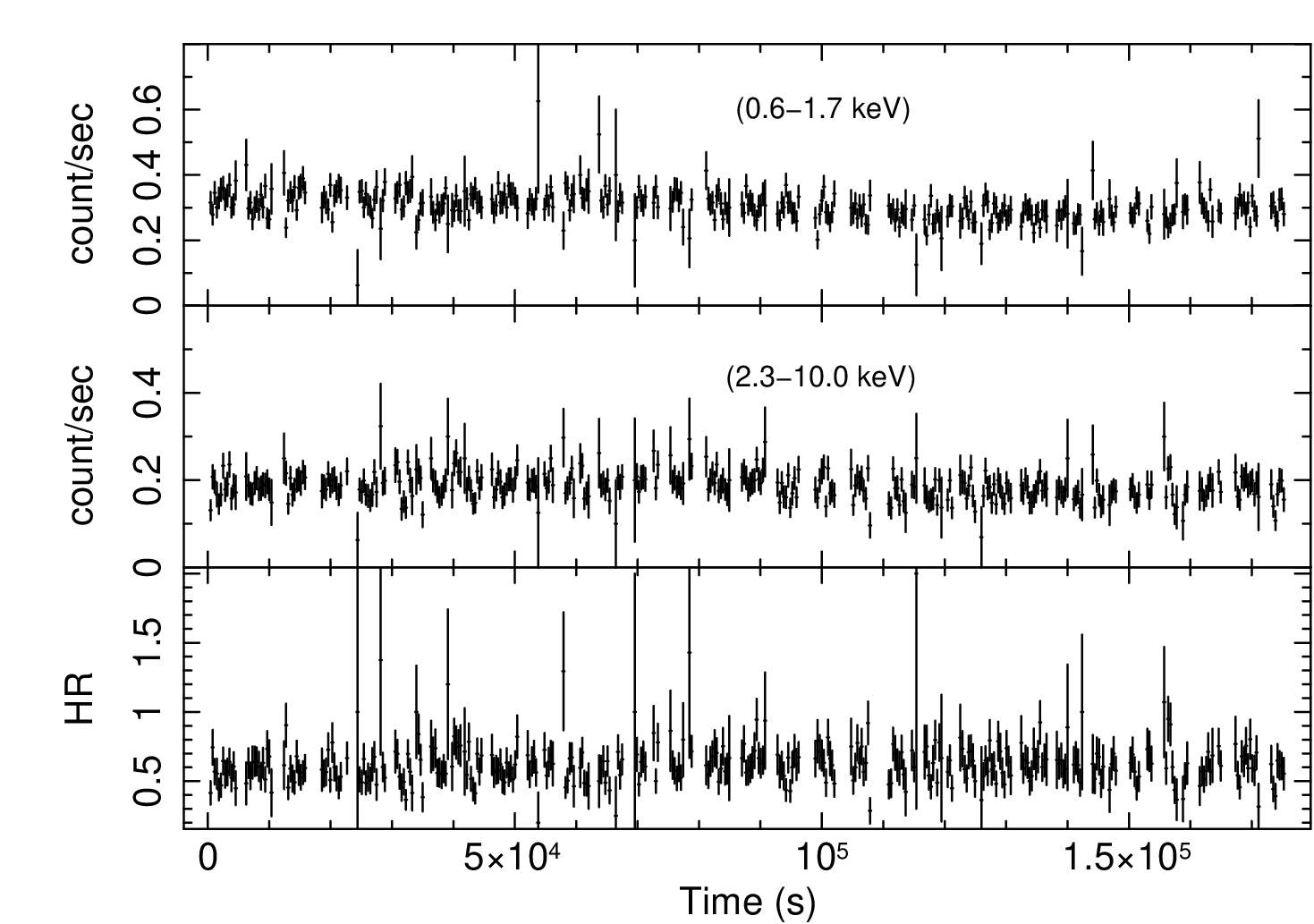}
\includegraphics[width=5.0cm,angle=-90]{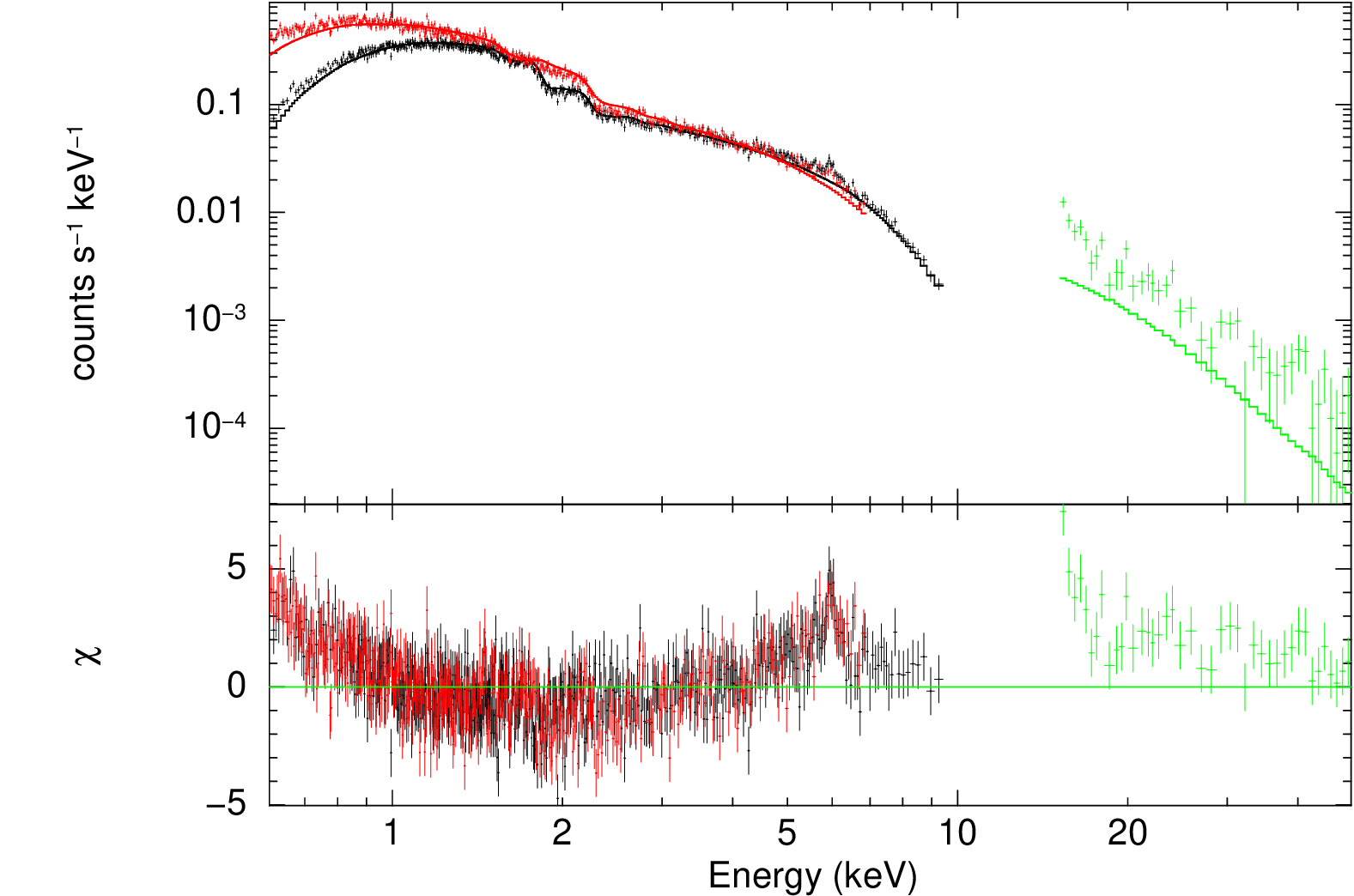} 
}\caption{ {\it Left:} Background subtracted XIS lightcurve of Mrk~205 in the $0.6-1.7 \kev$ energy band (top panel), $2.3-10 \kev$ (middle panel) and the hardness ratio (bottom panel) for the \suzaku{} observation of Mrk~205. Note that the hardness ratio does not vary significantly during the $101\ks$ observation.  {\it Right} Top panel: The $4.0-5.0\kev$ \suzaku{} spectra of Mrk~205 fitted with an absorbed powerlaw and the rest of the $0.6-50.0\kev$ dataset extrapolated. Bottom panel: The broadband residuals from the fit above, showing the presence of soft X-ray excess, an Fe line complex and a hard X-ray excess (at $E>10\kev$). The X-axis represents observed frame energy.} \label{fig:Suzaku_excess}

\end{figure*}

Figure \ref{fig:Suzaku_excess} right panel shows the extrapolation of a powerlaw model in the $4-5\kev$ energy band to the whole sensitive band pass of the \suzaku{} instruments. We clearly find the presence of a soft X-ray excess (at $E<2\kev$), an Fe line complex (at $6-7\kev$) and a hard X-ray excess ($E>10\kev$) in the residuals. Figure \ref{fig:Suzaku_excess} left panel shows the hardness ratio (HR) lightcurve of Mrk~205 for the \suzaku{} observation, where the HR is the ratio of the flux in the hard ($2.3-10\kev$) and soft ($0.6-1.7\kev$) energy band. We find that the HR does not show any significant variation during the observation. Hence, we used time averaged \suzaku{} spectra in this work. In this study, we used two sets of models to fit the broad band \suzaku{} as well as \xmm{} spectra. 1. The baseline phenomenological models and 2. The physical models, which we discuss below.

\subsection{The phenomenological models}\label{subsec:pheno}

We used a set of phenomenological models to describe the continuum as well as the discrete components in the X-ray spectra of Mrk~205. The sole purpose of this exercise is to identify the features that are present in the broad band spectra which can serve as a motivation for the use of specific physical models in the next section. The baseline phenomenological model consists of a neutral Galactic absorption (tbabs), a neutral absorption intrinsic to the host galaxy (ztbabs), a multiple blackbody component, which is used to model the soft X-ray excess \citep[diskbb,][]{1984PASJ...36..741M}, and the coronal emission described by a power law. For the discrete emission features, we used the {\it diskline} model \citep{1989MNRAS.238..729F} to describe the broad Fe line, and a Gaussian profile for the narrow Fe emission line.

We fixed the column density of {\it tbabs} to the Galactic value of $3\times 10^{20}\cmsqi$ \citep{1990ARA&A..28..215D}. We detected an intrinsic neutral absorption column for all the \xmm{} observations. Interestingly we did not detect any neutral absorption column excess of the Galactic column in the \suzaku{} spectra. An addition of the {\it ztbabs} model resulted no improvement in the fit ($\rm \dc/dof=2/1$) in the \suzaku{} spectra. Table \ref{Table:pheno} lists the best fit parameters and also the fit statistics achieved using the phenomenological models. We also list the improvement in statistics ($\rm \dc/dof$) for each spectra on addition of the discrete emission line models to assess the statistical significance of the model. An energy-independent multiplicative factor was used to account for the relative normalizations of different instruments of \suzaku{}. In {\it XSPEC} notation the baseline model reads as: {\tt constant$\times$ tbabs$\times$ztbabs$\times$(powerlaw+diskbb+diskline+\\Gaussian+pexrav)}. The same model was used for the EPIC-pn observations.

We detected a narrow $\fe$ emission line at $\sim 6.4\kev$ rest frame, for all of the \suzaku{} and \xmm{} observations. A Gaussian profile fit to the narrow line in \suzaku{} yielded an equivalent width of $60_{-11}^{+13}\ev$, with an improvement in the fit by $\rm \dc/dof=101/3$, which reflects a $>99.99\%$ confidence in the detection of the emission line \citep{1976ApJ...208..177L}. For \xmm{} observations a similar fit yielded equivalent width values of $49\pm 19\ev$, $35\pm15\ev$, $66\pm 14\ev$ and $10\pm3\ev$, with an improvement in the fit by $\dc=13, \, 26, \, 24$ and $32$ for xmm101, xmm201, xmm301 and xmm501 respectively (See Table \ref{Table:pheno}).

To account for the neutral reflection from distant neutral material, we used the model {\it pexrav} \citep{1995MNRAS.273..837M}, in the \suzaku{} spectra only, due to availability of data beyond $10\kev$. The {\it pexrav} incident powerlaw slope was tied to the primary powerlaw slope ($\Gamma$). The {\it pexrav} angle could not be constrained and hence was tied to the {\it diskline} inclination. The model {\it pexrav} in the \suzaku{} spectra improved the fit significantly with a $\dc \sim50$ and we obtained a best fit relative reflection fraction of $R=-1.69_{-0.22}^{+0.31}$.

Apart from a narrow emission line, a broad Fe emission line was also detected in the \suzaku{} spectra which was fitted with a {\it diskline} model, which improved the fit by $\rm \dc=37$ for three parameters of interest, with the rest frame line energy $E=6.36^{+0.31}_{-0.32}\kev$. The Fe abundance of the {\it diskline} model was fixed to that of the solar value. A similar fit with the diskline model for the \xmm{} observations resulted in relatively weak improvements in statistics, $\dc=17,\, 5, \, 0$ and $18$ for xmm101, xmm201, xmm301 and xmm501 respectively, clearly indicating that a broad emission line is not required for the observations xmm201 and xmm301. In the \suzaku{} spectra, the diskline inclination could be constrained to $22^{+4}_{-3}$ degrees, and this value was fixed for the \xmm{} observations, because we do not expect it to vary on human timescales.

\begin{figure*}
  \centering 

{
\includegraphics[width=8cm,angle=-90]{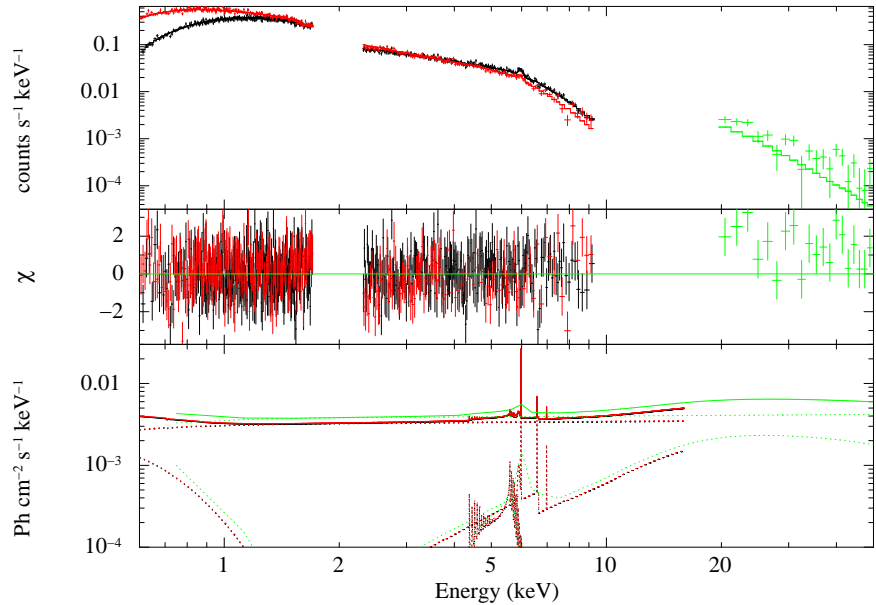} 
 }
	\caption{ The \suzaku{} spectra of Mrk~205 along with the best-fit baseline phenomenological models. Note that there is some excess visible in the {\it PIN} data after obtaining the best fit, which could be due to the relativistic disk reflection hump which is not modeled in the spectra. We have used a {\it diskline} profile to describe the broad $\fe$ emission line. See Section \ref{sec:analysis} for details. The X-axis represents observed frame energy.} \label{fig:suzaku_model}
\end{figure*}

 The baseline model gives a good description to the \suzaku{} as well as \xmm{} spectra (see Figure \ref{fig:suzaku_model}). The powerlaw photon index $\Gamma$ remains consistent ($\sim 1.9$) for all the observations within errors, similar to those obtained by previous studies \citep{2001A&A...365L.134R,2003MNRAS.343.1241P}. Two diskbb models were needed to fit the soft X-ray excess in the \xmm{} datasets with electron temperatures of $\sim 0.07\kev$ and $\sim 0.24\kev$. For \suzaku{} only one diskbb component was necessary with an electron temperature of $\sim 0.14\kev$.

 Previous study of the source Mrk~205 using \suzaku{} observation \citep{2016PASJ...68S..27I} has detected the presence of partially ionized cloud which could mimic the presence of a broad Fe K emission line \citep[see for e.g., ][]{2003PASJ...55..625I}. We fitted the \suzaku{} spectra with a partially covering ionized absorber model ({\it zxipcf}) and removed the {\it diskline} from the baseline model and found that the fit improves by $\dc=14$ compared to the baseline model fit described earlier. This implies a possiblity that the broad line is an artifact of a high column density partially covering ionised absorber. We investigate this matter further with physical models in the next section.

\subsection{The physical models}\label{subsec:phys}

 We used a set of physical models to describe the continuum and the discrete components in the \suzaku{} and \xmm{} X-ray spectra of Mrk~205. The narrow $\fe$ emission line along with the Compton reflection of hard X-ray photons off cold material is modeled with {\it MYTorus} \citep{2009MNRAS.397.1549M,2012MNRAS.423.3360Y}. The soft X-ray excess, the powerlaw, the broad Fe emission line (if any) and the reflection of hard X-ray photons off ionized accretion disk are modeled with {\it relxill} \citep[version 1.2.0, ][]{2014ApJ...782...76G}. The {\it MYTorus} inclination angle and the normalisation of the individual components were left free to vary. The photon index $\Gamma$ in {\it MYTorus} was tied with the primary power law component of the {\it relxill} model. The latest version of {\it relxill} model self consistently includes the relativistic broadening of the Fe K emission line along with the soft X-ray excess emission and the relativistic hard X-ray emission at $E>10\kev$. 
{The {\it relxill} model assumes a lamp-post geometry of scattering, whereby the hard X-ray emitter (corona) irradiates the accretion disk from the top, and the hard X-ray photons get Compton scattered from the accretion disk. In the model {\it relxill}, the accretion disk is assumed to be an optically thick slab of gas and the reflection emissivity from the accretion disk depends on its radius as a powerlaw $E(r) \propto r^{-q}$, where E is the emissivity of the gas due to reflection, and $q$ is the emissivity index. In Newtonian geometry, the emissivity index is $q=3$. In the general relativistic high gravity regime the inner emissivity profile steepens to higher values \citep{2011MNRAS.414.1269W}. The model {\it relxill} assumes that the transition from relativistic geometry to Newtonian geometry happens at a break radius $r_{\rm br}$. Thus in our fits, the emissivity index of the accretion disk at $r<r_{\rm br}$ have values ranging from $3<q<10$, while at $r>r_{\rm br}$ the index is fixed to $3$. Recent studies of the accretion disk emission using the model {\it relxill} \citep{2016MNRAS.456..554G,2018ApJ...855....3T,2018MNRAS.479.2464G,2018MNRAS.477.3711J} have found that the emissivity index in the relativistic regimes can take up values as large as $8-10$. For all the observations we have added a partially covering ionized absorption model {\it zxipcf} to account for any absorption along the line of sight.

 For {\it MYTorus} the equatorial column density and the inclination angle between the observer’s line of sight and the symmetry axis of the torus is allowed to vary freely. The inclination angles of the {\it relxill} and {\it MYTorus} models are untied and separate, because they are entirely two different reprocessing media and are treated as separate quantities. For the \xmm{} observation the inclination of MYTorus is not well constrained possibly due to non-coverage of $>10\kev$ spectrum.}

\begin{figure*}
  \centering

	\hbox{

		\includegraphics[width=6.5cm,angle=-90]{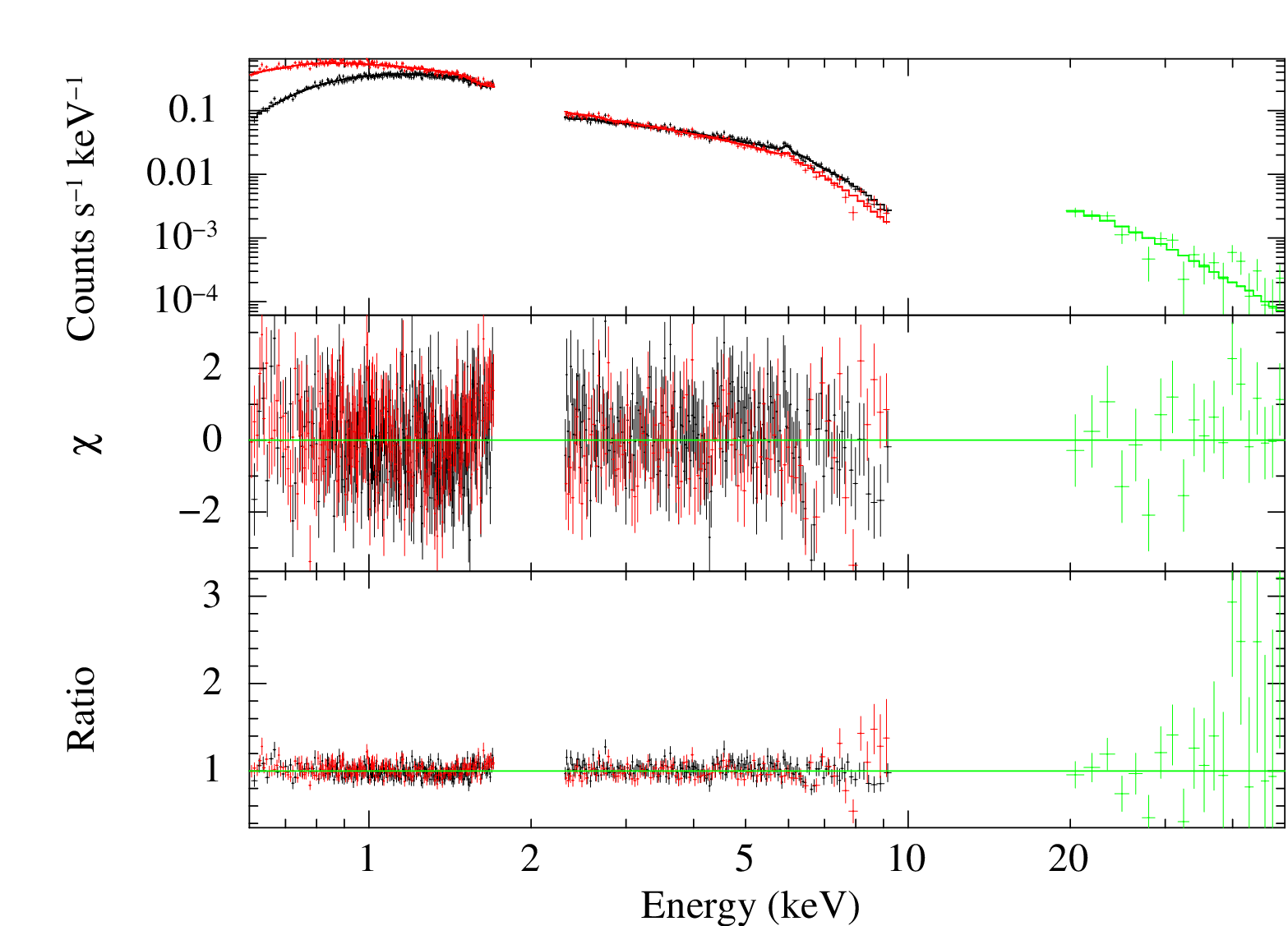}
		\includegraphics[width=6.5cm,angle=-90]{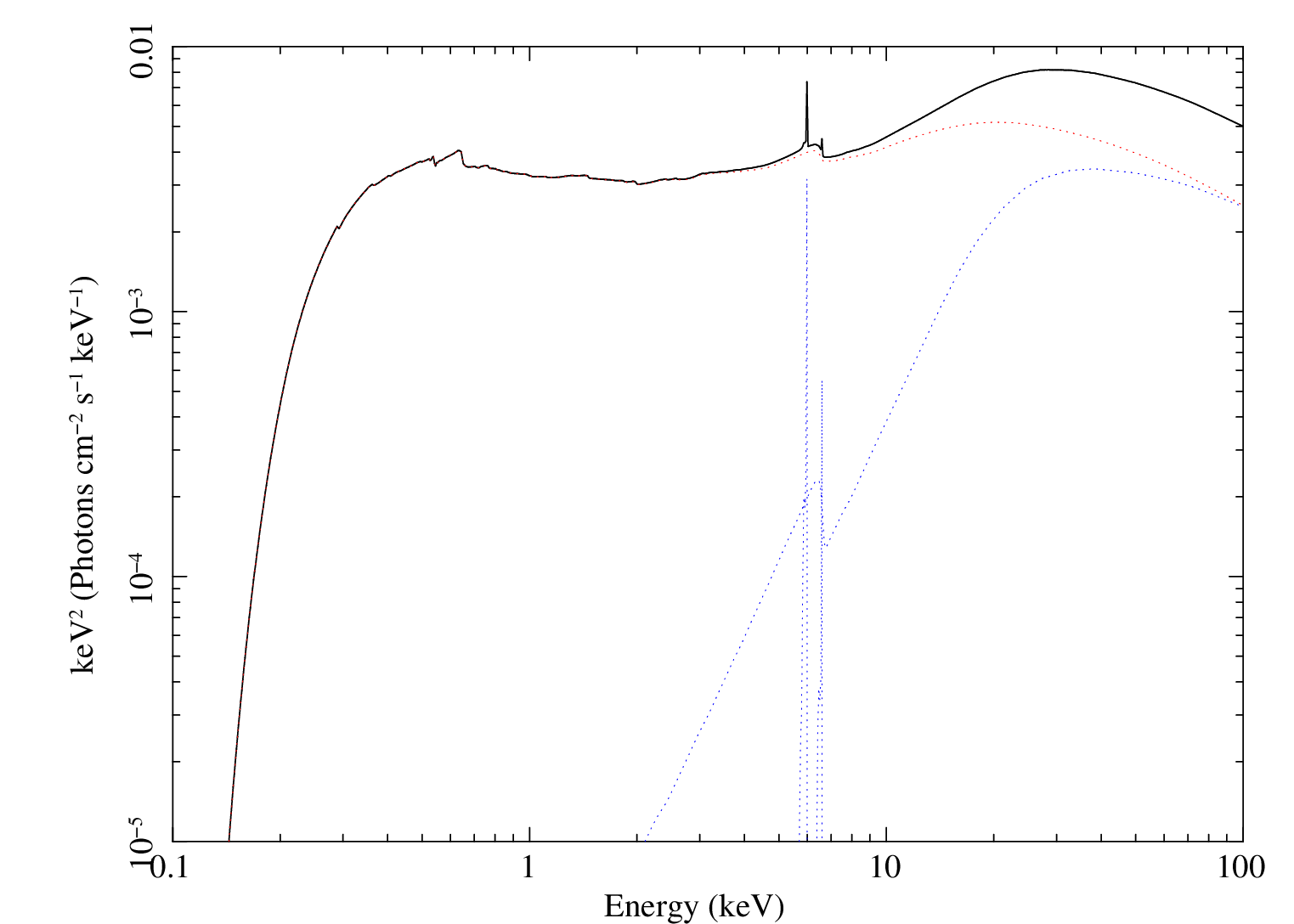} 
}
	\caption{{\it Left:} The \suzaku{} spectra of the source Mrk~205 with the best-fit physical models and residuals, as described in Section \ref{sec:analysis}. {\it Right:} The best fit physical models to the \suzaku{} spectra of Mrk~205. The {\it relxill} model describing simultaneously the soft X-ray excess, the broad $\fe$ emission line and the relativistic reflection hump in the hard X-rays is plotted in red dotted line. The {\it MYTorus} model which describes the narrow $\fe$, and Ni emission lines along with the Compton hump due to distant neutral reflection is plotted in blue dotted line. The final best fit model is plotted in black solid line. The X-axis represents observed frame energy.} \label{fig:suzakubestmodel}
\end{figure*}

\begin{figure*}
  \centering

	\hbox{

		\includegraphics[width=6.5cm,angle=-90]{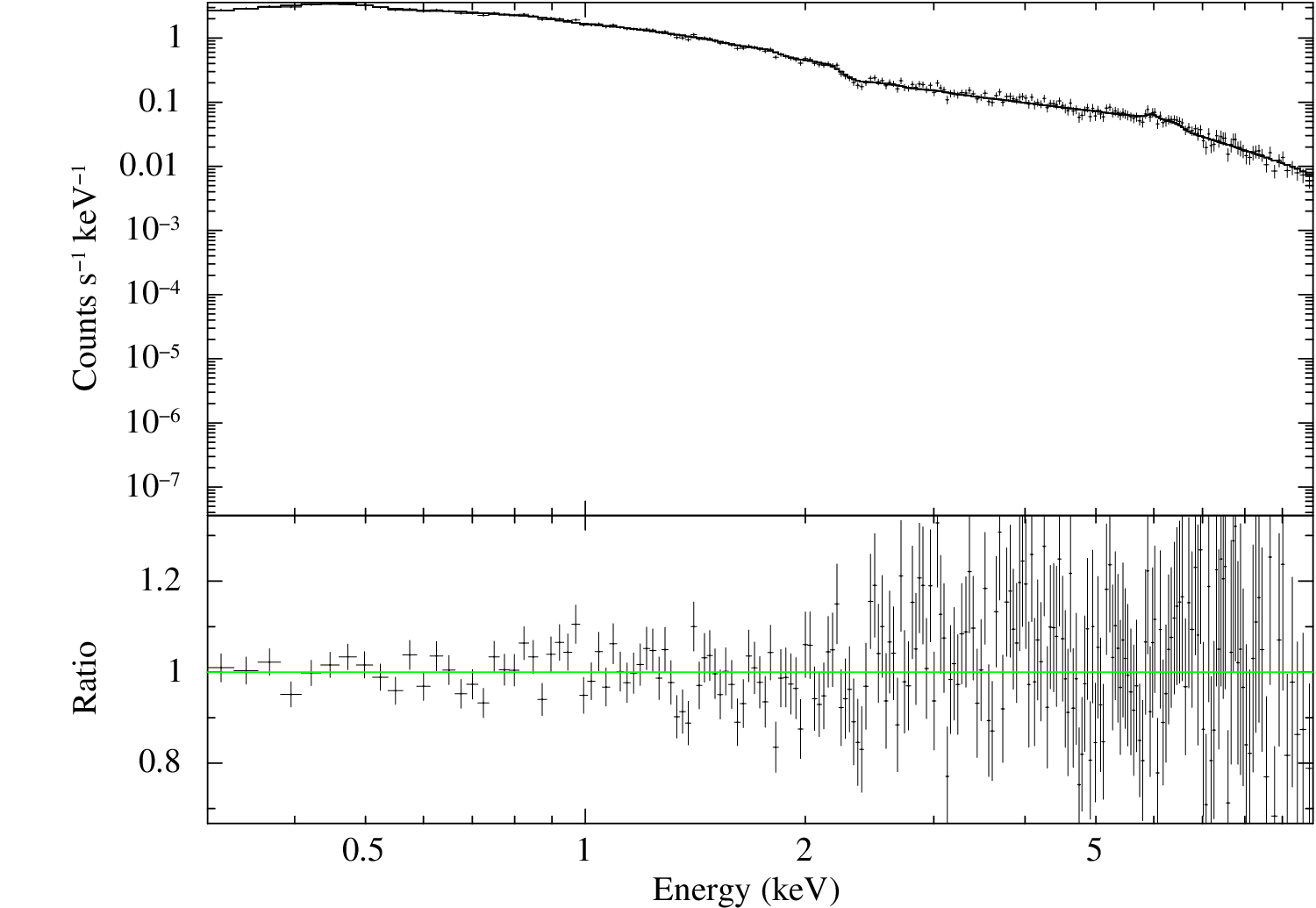}
		\includegraphics[width=6.5cm,angle=-90]{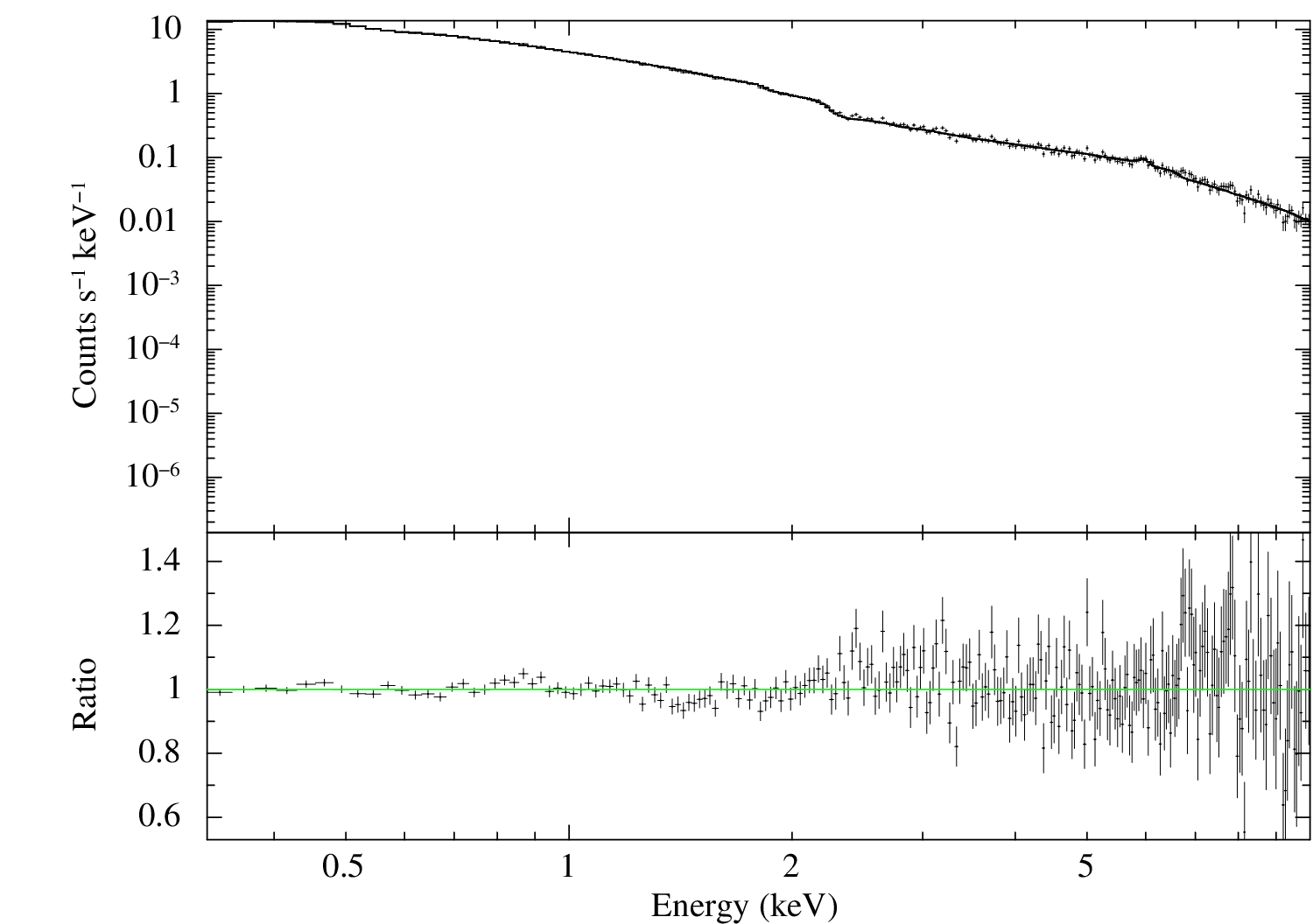} 
}

	\hbox{

		\includegraphics[width=6.5cm,angle=-90]{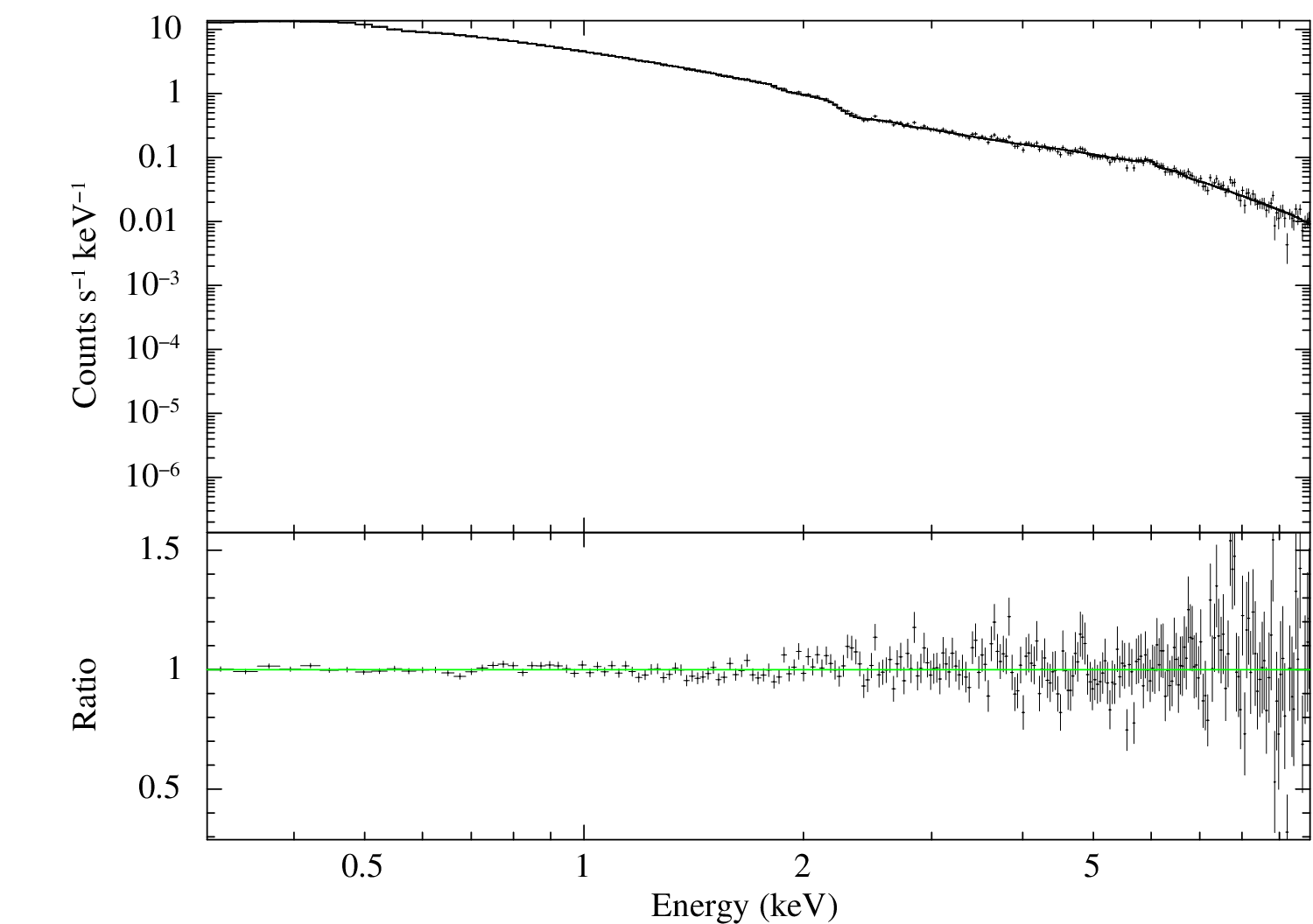}
		\includegraphics[width=6.5cm,angle=-90]{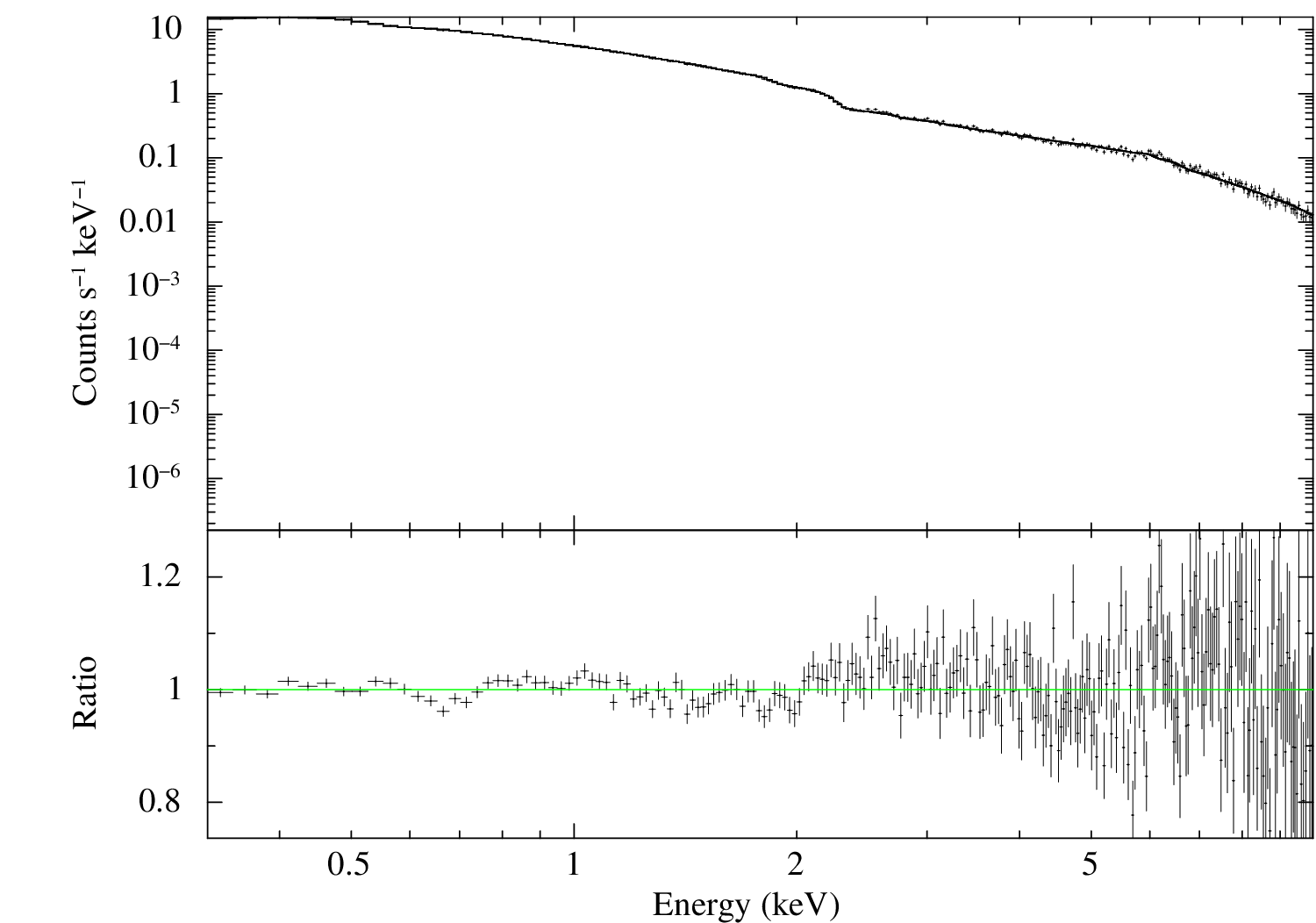} 
}

	\caption{{\it Top Left:} The \xmm{} spectra (id: xmm101) of the source Mrk~205 with the best-fit physical model and residuals, as described in Section \ref{sec:analysis}. {\it Top Right:} Same as left, except for the obsid: xmm201, {\it bottom Left:} Same as top left, except for the obsid: xmm301, {\it bottom Right:} Same as top right, except for obsid: xmm501} \label{fig:xmmbestmodel}
\end{figure*}

Table \ref{Table:relxill} lists the best fit model parameters along with the best fit statistics for \suzaku{} and \xmm{} observations. For the \suzaku{} spectra we have carried out two separate fits in order to ascertain if the data prefers a broad Fe line or not. In the first case we have kept the inner radius of the relativistic reflection model fixed to the inner circular stable orbit for a non rotating black hole ($R_{\rm in}=6R_{\rm G}$). In the second case, we fixed the inner radius to that of a maximally spinning black hole ($R_{\rm in}=1.23R_{\rm G}$). For the \xmm{} observations similar exercises have been done but we report only the Schwarzschild case because we do not find any difference in the quality of fit between the two scenarios. Figures \ref{fig:suzakubestmodel} and \ref{fig:xmmbestmodel} show the data, the best fit physical models and the residuals for the \suzaku{} and the \xmm{} observations. To test if the partially ionized absorption is needed by the data using the physical models, we removed the model ({\it zxipcf}) from the fit and tested the two cases of maximally spinning and non-spinning black hole. The fit statistic for the non-spinning case in the \suzaku{} observation is $845/735$, while for maximally spinning case is $836/735$ respectively. Both the fits are worse compared to those when the partially covering ionized absorber model is included, implying that the model is statistically required by the data. 

 We also tested for the case when the soft X-ray excess could be described by intrinsic disk-comptonization of the accretion disk photons emitted in the UV. The model {\tt optxagnf} \citep{2012MNRAS.420.1848D} assumes that the gravitational energy released in the accretion disk powers the disk emission in UV, the soft X-ray excess and the power law emission. The broadband model in {\it XSPEC} notation we used to fit the \suzaku{} spectra is {\tt (tbabs*zxipcf*(optxagnf+MYTorus))}. The fit resulted in a statistic $\cd=900/735$ (with positive residuals in the soft X-ray region) which is worse than the best fit model obtained with the baseline model above. This may indicate that the spectra needs a further disk reflection model to describe the soft excess. We added the model {\it relxill} to the current fit and the fit improved to $\cd=808/727$. The fit is comparable to that obtained with the physical model mentioned above. We have used only the Schwarzschild case for the reflection model here, as we have already checked that the data are insensitive to the spin of the black hole. We carried out the same exercise for the \xmm{} observations and found similar results. Table \ref{Table:optxagnf} lists the best fit parameters obtained using this model. We discuss these results in the next section.

%%%%%%%%%%%%%%%%%%%%%%%%%%%%%%%%%%%%%%%%%%%%%%%%%%%%%%%%%%%%  Results and discussion

\section{Results and discussion}\label{sec:results}

We have carried out an X-ray spectral analysis of the Seyfert 1 galaxy Mrk~205 using the observations from \suzaku{} and \xmm{} telescopes.  Previous studies by \citet{2001A&A...365L.134R} and \citet{2003MNRAS.343.1241P} could not affirmatively confirm the presence of broad $\fe$ emission line and the authors proposed that the higher ionization broad line could also be a blend of several higher ionization Fe K emission lines, mimicking a broad feature. In this study, for the first time, using broad band spectra ($0.5-30 \kev$) from \xmm{} and \suzaku{} and physical models such as {\it relxill}, we investigate the presence of a broad $\fe$ emission line in the source Mrk~205.

{The super massive black hole at the center of the galaxy Mrk~205 has a mass $10^{8.32\pm1.0}\times \msol$ \citep{2012MNRAS.422L...1T,2016MNRAS.462..511K}. The $2-10\kev$ unabsorbed luminosity of Mrk~205 is $L_{2-10\kev}=10^{43.8}\lunit$. Using the bolometric correction factor of $\kappa=20$ \citep{2007MNRAS.381.1235V}, we find that the bolometric luminosity of the AGN is $\lbol \sim 10^{45.10}\lunit$, which implies an Eddington rate of $\lambda_{\rm Edd}=0.03$. Below we discuss the main results in context to the origin of the narrow and broad Fe emission lines and the properties of the X-ray reprocessing media in the vicinity of the SMBH.}

\subsection{Origin of the narrow $\fe$ emission line}

The \suzaku{} X-ray spectra of Mrk~205 shows the presence of a narrow $\fe$ emission line with an equivalent width of $\sim 60\ev$, which is typical of nearby Seyfert galaxies. This fluorescent emission line is believed to arise when the hard X-ray photons from the corona get reflected from high column-density neutral material in the vicinity of the SMBH, possibly the torus. \citet{2014A&A...567A.142R} in a multiwavelength study of the narrow $\fe$ emission line found that the line originates from a neutral material possibly located in the molecular torus. The authors claim that most of the narrow $\fe$ emission is produced by the same material which is also responsible for emission in the mid-infrared. \citet{2011ApJ...727...19F} studied the X-ray spectra of a sample of AGN with \suzaku{}. They found a strong correlation between the equivalent width of the neutral $\fe$ emission line and the neutral absorption column density in the range $10^{23}-10^{25}\cmsqi$, indicating that the $\fe$ line is emitted by the Compton thick reprocessing medium around the SMBH. For absorption column densities below $10^{23}\cmsqi$ there was no dependence of the $\fe$ equivalent width on the column. In Mrk~205 we find that the {\it MYTorus} model, which models the narrow $\fe$ emission line simultaneously with the Compton hump (peaking at $\sim 20\kev$) has a best fit reflection column density of $\ge 2.15\times 10^{24}\cmsqi$. This result corroborates the fact that the neutral $\fe$ emission line of Mrk~205 arises from the Compton-thick torus.

\subsection{Is there a broad Fe line in the X-ray spectra?}

\citet{2001A&A...365L.134R} detected a broad $\fe$ emission line for the source Mrk~205 with \xmm{} spectra, however, the authors were skeptical about the nature of the emission line. The best fit rest frame line energy of the broad line measured was $6.7\kev$ which is not consistent with that of other broad line profiles measured in Seyfert-1 galaxies \citep{1997ApJ...477..602N}, where the line peaks at around $6.4\kev$ and the bulk of the line flux is redshifted below this energy. The authors also discussed the possibility of a large inclination angle of the disk $\sim 75-90$ degrees, but it would mean that the source is nearly edge-on, defying the optical classification of it being a Seyfert 1 galaxy. Another possiblity could be that the inner accretion disk being too highly ionized, the red wing of the broad $\fe$ emission line is suppressed. The authors also proposed an alternative scenario whereby the apparent broad emission line is due to the blend of several high ionization Fe lines near $6.7 \kev$. It is common to find ionized emission lines in Seyfert 2 spectra arising from warm electron scattering regions, but the equivalent widths of such lines are much smaller in case of Seyfert 1 galaxies. More recent comprehensive sample study on broad $\fe$ emission lines \citep{2010A&A...524A..50D} have detected broad $\fe$ emission line peaking at $\sim 6.4\kev$ with a substantial flux in the red wing of the profile.

For the observation xmm101, same as the one studied by \citet{2001A&A...365L.134R} and \citet{2003MNRAS.343.1241P}, we detect an Fe emission line at $6.84_{-0.14}^{+0.15}$, consistent with the previous studies, using simple Gaussian fits. To test the presence of a broad emission line arising out of inner-accretion-disk reflection of a spinning black hole we used the relativistic reflection models. We created two sets of models to fit the \suzaku{} and \xmm{} observations. One with the inner circular stable orbit fixed to value as obtained for a non rotating black hole (i.e, $R_{\rm in}=6R_{\rm G}$). In the second case, we fixed the inner radius to that of a maximally spinning black hole ($R_{\rm in}=1.23R_{\rm G}$). Table \ref{Table:relxill} shows the best fit parameters for the physical model, and we find that both the models result in equally good fit. The results indicate two possible scenarios: 1. Either the broad Fe line is indeed not present or 2. The data are not sensitive enough to detect them.

\subsection{Is the disk truncated for this source?}

As discussed above, the \suzaku{} and \xmm{} observations pointed out that the data cannot distinguish between the two scenarios: a rapidly spinning and a non-spinning black hole. This may imply that we are viewing a truncated accretion disk whose inner radius lie further out ($>6R_{\rm G}$). In the \suzaku{} observation we have carried out a test to check how far the inner radius of the disk extends in order to produce simultaneously the soft X-ray excess and the hard excess. We set the inner radius in the Schwarzschild case to $>20R_{\rm G}$, and froze the inner emissivity index to the Newtonian value, $q1=3$. We measured the range of the inner radius and found an upper limit of $<12R_{\rm G}$. We therefore conclude that the inner radius of the accretion disk may have been truncated at a larger distance in the range $6R_{\rm G}<R<12R_{\rm G}$. We note however, that the truncation radius is not very significantly large, implying that the relativistic effects could still be present and the disk reflection model is required by the data. This is also corroborated by the fact that we did not get a good fit when we used disk Comptonization model ({\it Optxagnf}) to fit the data, and the fit improved radically upon the addition of a disk reflection model ({\it relxill}). The truncation of the disk is also supported by the fact that the Eddington rate of the source is low ($\lambda_{\rm Edd}=0.03$), implying that the black hole is not feeding efficiently.  

%We should however note as a caveat, that estimating the disk inner radius using reflection models, in the cases where the disk has been truncated at a larger radius, has its own issues \citep[see for e.g., ][]{2014MNRAS.439.2307F}. 

%With the physical models, we find that the broad $\fe$ emission line may originate from the inner accretion disk which is highly ionized. The model {\it relxill} could describe the broad line simultaneously with the soft-excess, and the hard excess, indicating that all of these originate from the same region of the inner accretion disk (see Figure \ref{fig:suzakubestmodel}). The inner radius of the accretion disk is measured to be $<1.3\, r_{\rm G}$ and we could constrain the black hole spin $a=0.96_{-0.01}^{+0.03}$, which implies that the SMBH at the centre of Mrk~205 could be spinning very rapidly.  

%%%%%%%%%%%%%%%%%%%%%%%%%%%%%%%%%%%%%%     Conclusions

\section{Conclusions}\label{sec:conclusions}

We have carried out an extensive X-ray spectral analysis of the local Seyfert-1 galaxy Mrk~205, investigating the anomalous hard X-ray emission of the source Mrk~205 as shown by previous studies. We used the broad band spectra from \xmm{} and \suzaku{} for our investigation. We list the main conclusions below:\\

\begin{itemize}

	\item{The X-ray spectra of Mrk~205 is typical of local Seyfert galaxies, with power law photon index $\Gamma\sim1.90$, a soft-excess, an Fe line complex and a hard X-ray excess. The soft excess possibly arise from the ionized accretion disk, as it could be described simultaneously with the hard X-ray excess. A fit with intrinsic disk comptonization model ({\it optxagnf}) did not result in a good fit.}

	\item {The origin of the narrow $\fe$ emission line is possibly due to the reflection of hard X-ray photons off the neutral molecular torus with a column density of $\ge 2.15\times 10^{24}\cmsqi$, implying a Compton thick reflector. }

	\item We detect a partially covering ionized absorption with ionization parameter $\log(\xi/\xiunit)=1.9_{-0.5}^{+0.1}$, column density $\nh=(5.6_{-1.9}^{+2.0})\times 10^{22}\cmsqi$ and a covering factor of $0.22_{-0.06}^{+0.09}$

	\item{We conclude that we cannot affirmatively confirm the presence of a broad line. The data quality is not sensitive enough to detect the broad line. However, we confirm that disk reflection component is statistically required by all the spectra.}

	\item{We found that the accretion disk may be truncated at a larger distance from the SMBH, $6R_{\rm G}<R<12R_{\rm G}$, as obtained from broad band \suzaku{} observation, confirming the fact we do not detect any broad Fe emission line. This is supported by the fact that the Eddington rate of the source is low ($\lambda_{\rm Edd}=0.03$).}

\end{itemize}

%%%%%%%%%%%%%%%%%%%%%%%%%%%%%%%%%%%%%%%      Table for X-ray data.

%\begin{sidewaystable}
\begin{table*}

{\footnotesize
\centering
  \caption{The X-ray observations of Mrk~205. \label{Table:obs}}
  \begin{tabular}{llllllllllllllll} \hline\hline 

X-ray		& observation	&Short	&Date of obs	& Net exposure	\\
Satellite	&id		&id	&		&	&\\ \hline 

	  \xmm{}		&{0124110101}	&xmm101	&07-05-2000	& $15\ks$	&\\
		&{0401240201}	&xmm201	&18-10-2006	& $28 \ks$			&\\
		&{0401240301}	&xmm301	&20-10-2006	& $29\ks$			&\\
		&{0401240501}	&xmm501	&22-10-2006	& $43 \ks$			&\\

	  \suzaku{}	&705062010		&	&22-05-2010	&$101\ks$ &	\\ 
	  \nustar{}$^1$&60160490002		&	&20-06-2017	&$20\ks$	\\

\hline 
	
\end{tabular}  

	{$^1$ Due to insufficient signal to noise ratio of the \nustar{} observation in the FeK band, the data were not used for analysis in this paper. See Appendix \ref{append:nustar} for details of the spectra and fit.}\\
	%{$^1$ The \chandra{} and the \nustar{} observations were not included in this work due to lack of adequate signal to noise ratio required to carry out spectral analysis. Appendix A elaborately discusses the \chandra{} and \nustar{} spectra of Mrk~205.}

}
\end{table*}
%\end{sidewaystable}

%%%%%######################

\clearpage

%%%%%######################

%%%%%######################

%%%%%######################

%%%%%%%%%%%%%%%%%%%%%%%%%%%    XMM fits...

%%%%%%%%%%%%%%%%%%%%%%%%%%%%%%%%%%%%%%%%%%%%%%%%%    Table 2  

\begin{table*}

\centering
%{\large

  \caption{The best fit parameters of the baseline phenomenological models for the \suzaku{} and \xmm{} observations of Mrk~205.  \label{Table:pheno}}
{\renewcommand{\arraystretch}{1.5}
\setlength{\tabcolsep}{1.5pt}
% \scalebox{0.9}{
\begin{tabular}{ccccccc} \hline\hline
	
Models 		& Parameter 				& Suzaku 		& xmm101 			& xmm201 		& xmm301 			& xmm501  \\  \hline 

Gal. abs.  	& $\nh \,(\times 10^{20}\, \cmsqi)$ 	& $ 3.0$ (f)     	& $ 3.0$ (f)     		& $ 3.0$ (f)     	& $ 3.0$ (f)     		& $ 3.0$ (f) \\

ztbabs   	&   $\nh$ ($\times 10^{20}\cmsqi$)      & ---   	 	& $8^{+2.0}_{-1.0}$ 		& $4^{+2.0}_{-1.0}$ 	& $-$ 				& $2^{+1.0}_{-1.0}$ \\

 powerlaw 	& $\Gamma$         			& $1.97^{+0.01}_{-0.01}$ & $1.72^{+0.05}_{-0.03}$ 	& $1.94^{+0.05}_{-0.05}$ & $1.90^{+0.02}_{-0.06}$ 	&  $1.94^{+0.03}_{-0.02}$ \\
         	 & norm ($10^{-3}$) 			& $3.19^{+0.04}_{-0.02}$ & $1.32^{+0.05}_{-0.05}$ 	& $3.12^{+0.25}_{-0.19}$ & $2.88^{+0.18}_{-0.18}$ 	&  $4.19^{+0.13}_{-0.13}$ \\

diskbb(1) 	& $T_{in}$ (keV)   			& ---                   & $0.07^{+0.01}_{-0.01}$ 	& $0.08^{+0.01}_{-0.01}$ & $0.11^{+0.01}_{-0.01}$ 	&  $0.09^{+0.01}_{-0.01}$ \\
          	& norm ($10^{+4}$) 			& ---                   & $2.42^{+2.80}_{-0.50}$ 	& $2.72^{+2.10}_{-1.50}$ & $0.41^{+0.60}_{-0.40}$ 	&  $1.56^{+0.71}_{-0.49}$ \\

diskbb(2) 	& $T_{in}$ (keV)   			& $0.12^{+0.01}_{-0.01}$ & $0.23^{+0.01}_{-0.02}$ 	& $0.25^{+0.01}_{-0.01}$ & $0.32^{+0.03}_{-0.03}$ 	&  $0.27^{+0.02}_{-0.03}$ \\
          	& norm             			& $2502^{+541}_{-356}$   & $37^{+11}_{-15}$       	& $78^{+49}_{-38}$       & $18^{+19}_{-11}$ 		&  $45^{+16}_{-12}$ \\

diskline  	& E($\kev$)        			& $6.36^{+0.31}_{-0.32}$ & $6.84^{+0.15}_{-0.14}$ 	& $-$ 			& $-$ 				&  $6.36^{+0.31}_{-0.32}$ \\
          	& $\beta$          			& $ <-5.97 $             & $ <-2.58 $              	& $-$			& $-$ 	        		&  $ <-2.51$  \\
		& $ R_{\rm in} (r_g)$    		& $ <9.79  $             & $ <578$     	& $-$               	& $-$		 				&   $ <32 $ \\
          	& Incl in($^\circ$)			& $ 22^{+4}_{-3}$        & $ <58 $                	& $-$               	& $-$	          		&   $ <30 $ \\
          	& norm ($10^{-5}$) 			& $1.33^{+0.48}_{-0.45}$ & $1.71^{+0.60}_{-0.60}$ 	& $-$ 			& $-$			  	&  $1.55^{+0.65}_{-0.44}$ \\
		&$\rm ^A$$\rm \dc/dof$			&$37/5$			 &$17/5$			&$5/5$			&$0/5$				&$18/5$	\\

Narrow-Gauss	&E($\kev$)        			& $6.38^{+0.15}_{-0.29}$ & $6.36^{+0.09}_{-0.14}$ 	& $6.36^{+0.25}_{-0.19}$ & $6.39^{+0.58}_{-0.80}$ 	&  $6.58^{+0.49}_{-0.10}$ \\
		&norm ($10^{-6}$) 			& $6.28^{+0.53}_{-0.73}$ & $3.81^{+2.80}_{-2.70}$ 	& $4.03^{+1.44}_{-1.51}$ & $6.88^{+0.33}_{-0.31}$  	&  $1.53^{+0.65}_{-0.44}$ \\

		&$\rm ^A$$\rm \dc/dof$			&$23/3$			&$13/3$				&$26/3$			&$24/3$				&$32/3$		\\

{ pexrav}$^B$    	& $R$              		& $-1.69_{-0.22}^{+0.31}$ & ---                   & --- 		        & ---             &   --- \\
			& Incl             		& $22^{\rm C}$            & ---                   & --- 		        & ---             &   --- \\ \hline

			$\cd$     &                  				& $ 834/737 $            	& $ 221/207 $            & $ 309/237 $            & $ 330/242 $      &       $ 342/248 $ \\\hline
\end{tabular}  

{$\rm ^A$ The $\dc$ improvement in statistics upon addition of the corresponding discrete component. }\\
{$\rm ^B$ The model {\it pexrav} was used only for \suzaku{} observation as it had broad band spectra necessary for constraining the parameters.}\\
{$\rm ^C$ The inclination angle of {\it pexrav} is tied to the inclination angle of the diskline as it could not be constrained independently.}

}

\end{table*}

%%%%%%%%%%%%%%%%%%%%%%%%%%%%%%%%%%%%%%%%%%%%%%%%

\begin{table*}
\footnotesize
%\large
\centering
  \caption{Best fit parameters for observations of Mrk~205 with the physical models. \label{Table:relxill}}

  \begin{tabular}{llllllll} \hline
Component  & parameter                & \suzaku{}     	       &                                & xmm101 		& xmm201 		& xm301 		& xmm501  \\
           &                          & Without Spin           & With Spin                      & Without Spin          & Without Spin          & Without Spin          & Without Spin \\\hline

Gal. abs.  & $\nh (10^{20} cm^{-2})$  & $ 3.0$ (f)     	       & $ 3.0$ (f)                     & $ 3.0$ (f)     	& $ 3.0$ (f)     	& $ 3.0$ (f)            & $ 3.0$ (f) \\

zxipcf    &   $\nh$ ($\times 10^{22}\cmsqi$) & $5.6^{+2.0}_{-1.9}$ & $5.1^{+2.0}_{-2.6}$ 	&  $<0.06$              &$48.3^{+22.0}_{-29.0}$ & $16.5^{+30.5}_{-6.2}$  & $23.3^{+25.3}_{-4.4}$\\

          &  $log \xi(\ergs)$                & $1.9^{+0.1}_{-0.5}$ & $1.9^{+1.0}_{-1.0}$        & $-1.6^{+0.2}_{-0.4}$  & $<0.52$               & $<0.03$                & $<0.53$\\

          &  $Cvr_{frac}$                    & $0.22^{+0.09}_{-0.06}$ & $0.22^{+0.11}_{-0.07}$  & $0.65^{+0.21}_{-0.26}$& $0.34^{+0.17}_{-0.11}$& $0.28^{+0.11}_{-0.15}$ & $0.22^{+0.13}_{-0.08}$ \\\hline

{\it relxill }  &  $A_{Fe}$                  & $1$(f)  	        & $1$(f)                        & $1$(f)                & $1$(f)    	        & $1$(f)                 & $1$(f)  \\

           	&  $log\xi (\xiunit)$       & $1.23^{+0.28}_{-0.26}$  & $1.32^{+0.79}_{-0.79}$ 	& $1.29^{+0.10}_{-0.25}$& $0.30^{+0.06}_{-0.23}$& $0.40^{+0.34}_{-0.22}$ & $0.38^{+0.34}_{-0.20}$\\ 

           	& $ \Gamma $               & $2.19^{+0.05}_{-0.05}$& $2.19^{+0.07}_{-0.05}$ 	&$2.13^{+0.04}_{-0.04}$ &$2.39^{+0.02}_{-0.02}$ &$2.36^{+0.02}_{-0.02}$  &$2.27^{+0.01}_{-0.01}$\\

           	&  $n_{rel}(10^{-5})^a$    & $5.52^{+0.16}_{-0.30}$  & $5.30^{+0.42}_{-0.62}$   & $2.50^{+0.03}_{-0.04}$& $8.55^{+1.33}_{-1.53}$& $7.90^{+0.69}_{-0.33}$ & $8.78^{+1.03}_{-0.36}$ \\
	   
           	&   $ q1$                  & $6$(pegged)   	& $3.3^{+0.7}_{-1.9}$   	& $6$(pegged)  & $6$(pegged)   & $6$(pegged)  & $6$(pegged) \\
		&   $ q2$                  & $3$(f)   	& $3$(f)		   	        & $3$(f)  & $3$(f)   & $3$(f)  & $3$(f) \\
 
		&   $ a$                   & $0$ (f) 	& $0.99$ (f)                            & $0$ (f) & $0$ (f)  & $0$ (f) & $0$ (f)\\

           	&   $R(refl frac) $        & $1.0^{+0.3}_{-0.3}$& $1.5^{+0.8}_{-0.8}$ 	        & $0.5^{+0.2}_{-0.2}$   & $0.73^{+0.25}_{-0.17}$& $0.50^{+0.22}_{-0.20}$  & $0.51^{+0.24}_{-0.12}$  \\

           	&   $ R_{in}(r_{g})$       & $6$(f)  	& $1.3$(f)                              & $6$(f)  & $6$(f)   & $6$(f)  & $6$(f)  \\

		&   $ R_{br}(r_{g})$       & $10$(f)         	& $10$(f)         	        & $10$(f) & $10$(f)  & $10$(f) & $10$(f)  \\

           	&   $ R_{out}(r_{g})$      & $400$ (f)      	& $400$ (f)      	        & $400$(f)& $400$(f) & $400$(f)& $400$(f)    \\

		&   $i(degree) $           & $31^{+5}_{-6}$	& $32^{+11}_{-8}$ 	        & $31$(f)      & $31$(f)         & $31$(f)               & $31$(f)\\ \hline

	  {\it MYTorusL}  &   $i(degree) $           & $<48$                 & $48$(f)   		        & $48$(f)   & $48$(f)    & $48$(f)   & $48$(f)\\

		&  norm ($10^{-3}$)        & $7.4^{+2.4}_{-2.2}$   & $8.2^{+2.6}_{-2.2}$ 	& $<6.2$                & $12.5^{+7.6}_{-7.7}$   & $7.4^{+5.9}_{-5.3}$ & $<7.4$\\

{\it MYTorusS } &  NH($10^{24} cm^{-2}$)   & $>2.15$         	& $>5.80$         	        & $10.0$(*)             &$10.0$(*)               &$10.0$(*)              &$10.0$(*)\\

		&  norm ($10^{-3}$)        & $3.30^{+2.50}_{-0.22}$& $3.16^{+2.51}_{-0.30}$ 	& $35.8^{+21.1}_{-3.3}$ &$87.3^{+4.4}_{-40.4}$   &$32.0^{+47.4}_{-13.8}$ &$58.7^{+22.7}_{-7.1}$\\\hline
		
	   	& $\cd $                   & $818/732$      	& $816/732$    		        & $241/206$             & $334/237$              & $322/241$             & $399/248$\\\hline 
\end{tabular} \\ 
Notes: (f) indicates a frozen parameter. (*) indicates parameters are not constrained.\\
	(a) $n_{rel}$ reperesent normalization for the model {\it relxill} %where $n_{pl}$ has the unit as photons $\kev^{-1} cm^{-2}s^{-1}$. 
%(b) $f_{PL}$ and $f_{REF}$ reperesent the $0.6-10\kev$ flux in the powerlaw and reflection components in  units of $10^{-11}$ $\funit$.
\end{table*}

%%%%%%%%%%%%%%%%%%%%%%%%%%%%%%%% Optxagnf+relxill+mytorus best-fit parameters %%%%%%%%%%%%%%%

\begin{table*}
\footnotesize
%\large
\centering
  \caption{Best fit parameters for observations of Mrk~205 with the physical models. \label{Table:optxagnf}}

  \begin{tabular}{llllllll} \hline
Component  & parameter                & \suzaku{}     	           & xmm101 		& xmm201 		& xm301 		& xmm501  \\
           &                          & Without Spin               & Without Spin       & Without Spin          & Without Spin          & Without Spin \\\hline

Gal. abs.  & $\nh (10^{20} cm^{-2})$  & $ 2.8$ (f)     	           & $ 2.8$ (f)     	& $ 2.8$ (f)     	& $ 2.8$ (f)            & $ 2.8$ (f) \\

zxipcf    &   $\nh$ ($\times 10^{22}\cmsqi$) & $4.9^{+2.1}_{-2.9}$ & $<5.1$ 	        &  $<0.06$              &$<2.08$                & $>4.7$  \\

          &  $log \xi(\ergs)$                & $1.9^{+0.2}_{-0.5}$ & $-1.4^{+2.0}_{-0.9}$& $<1.6$               & $<-0.52$              & $>1.5$               \\

          &  $Cvr_{frac}$                    & $<0.30$             &$0.28^{+0.11}_{-0.07}$& $<0.20$             & $<0.4$                & $>0.6$ \\\hline

optxagnf   & $ M_{BH}^a$            &$2.1$(f)                 &$2.1$(f)                &$2.1$(f)                &$2.1$(f)               &$2.1$(f)\\
           & $d{\rm~(Mpc}) $        &$308$(f)                 &$308$(f)                &$308$(f)                &$308$(f)               &$308$(f)\\
           &  $(\frac{L}{L_{E}})$   &$1.1^{+0.1}_{-0.7}$      &$0.1(*)$                &$0.4^{+0.7}_{-0.3}$     &$0.4^{+0.3}_{-0.3}$    &$0.11^{+0.2}_{-0.10}$\\ 
           &  $ kT_{e} (\kev)$      &$0.32^{+0.02}_{-0.02}$   &$0.26(*)$               &$0.48^{+0.12}_{-0.20}$  &$0.51^{+0.52}_{-0.21}$ &$0.51^{+0.15}_{-0.09}$\\ 
           &  $ \tau $              &$>5.5$                   &$9.9(*)$                &$>6.2$                  &$>4.3$                 &$9.4^{+1.6}_{-2.4}$\\
           &  $ r_{cor}(r_{g})$     &$9.9^{+6.8}_{-1.4}$      &$ 6.1(*) $              &$9.3^{+8.8}_{-1.4}$    &$9.0^{+3.8}_{-2.7}$    &$16$(*)\\
           &  $ a $                 &$0 $(f)                  &$ 0$(f)                 &$ 0$(f)                 &$ 0$(f)                &$0$(f) \\
           &  $ f_{pl}$             &$0 $ (f)                 &$ 0$ (f)                &$ 0$ (f)                &$ 0$ (f)               &$ 0$ (f)\\
           &  $ \Gamma $            &$2.04$(f)                &$ 1.93$(f)              &$2.15$(f)               &$2.16$(f)              &$2.10$(f) \\

{\it relxill }  &  $A_{Fe}$                  & $1$(f)  	             & $1$(f)                   & $1$(f)                & $1$(f)    	        & $1$(f)  \\

           	&  $log\xi (\xiunit)$       & $0.45^{+0.41}_{-0.26}$ & $3.01^{+0.11}_{-0.39}$ 	& $0.70^{+0.40}_{-0.15}$& $0.70^{+0.26}_{-0.23}$& $0.70^{+0.16}_{-0.22}$ \\ 

           	& $ \Gamma $               & $2.04^{+0.05}_{-0.05}$  & $1.93^{+0.06}_{-0.05}$ 	&$2.15^{+0.16}_{-0.12}$ &$2.16^{+0.12}_{-0.09}$ &$2.11^{+0.02}_{-0.02}$  \\

           	&  $n_{rel}(10^{-5})^b$    & $5.18^{+0.16}_{-0.30}$  & $2.27^{+0.42}_{-0.62}$   & $5.18^{+0.02}_{-0.04}$& $4.89^{+1.82}_{-1.53}$& $6.48^{+0.09}_{-0.02}$  \\
	   
           	&   $ q1$                  & $3.5^{+2.7}_{-1.9}$     & $6$(pegged)   	        & $4.6^{+4.5}_{-1.9}$  & $>4.1$   & $>3.9$   \\
 
		&   $ a$                   & $0$ (f) 	             & $0$ (f)                  & $0$ (f)               & $0$ (f)               & $0$ (f) \\

           	&   $R(refl frac) $        & $0.3^{+0.3}_{-0.3}$     & $0.3^{+0.2}_{-0.2}$ 	& $0.5^{+0.3}_{-0.2}$   & $0.5^{+0.2}_{-0.2}$   & $0.5^{+0.1}_{-0.1}$   \\

           	&   $ R_{in}(r_{g})$       & $6$(f)  	             & $6$(f)                   & $6$(f)                & $6$(f)                & $6$(f)  \\

		&   $ R_{br}(r_{g})$       & $12^{+0.3}_{-0.3}$      & $>10$(f)         	& $19$(*)               & $15^{+9.0}_{-8.3}$    & $<24$(f)  \\

           	&   $ R_{out}(r_{g})$      & $400$ (f)      	     & $400$ (f)      	        & $400$(f)              & $400$(f)              & $400$(f)   \\

           	&   $i(degree) $           & $4^{+7}_{-6}$	     & $18^{+9}_{-8}$ 	        & $19^{+10}_{-10}$      & $13^{+11}_{-6}$       & $20$(pegged)\\ \hline

{\it MYTorusL}  &   $i(degree) $           & $<60$                   & $<60$   		        & $<60$                 & $<60$                 & $<60$   \\

		&  norm ($10^{-3}$)        & $2.4^{+2.4}_{-2.2}$     & $2.4$(f) 	        & $2.4$(f)              & $2.4$(f)              & $2.4$(f) \\

{\it MYTorusS } &  NH($10^{24} cm^{-2}$)   & $>2.15$         	     & $10$(pegged)         	& $10$(pegged)          &$10$(pegged)           &$10$(pegged)\\

		&  norm ($10^{-3}$)        & $0.30^{+2.50}_{-0.22}$  & $0.30$(f) 	        & $0.30$(f)             &$0.30$(f)              &$0.30$(f) \\\hline
		
	   	& $\cd $                   & $805/727$      	     & $240/204$    		& $325/235$             & $310/239$             & $405/246$\\\hline 
\end{tabular} \\ 
Notes: (f) indicates a frozen parameter. (*) indicates parameters are not constrained. (a):in units of $10^7\rm M\odot$;\\
	(b) $n_{rel}$ reperesent normalization for the model {\it relxill} %where $n_{pl}$ has the unit as photons $\kev^{-1} cm^{-2}s^{-1}$. 
%(b) $f_{PL}$ and $f_{REF}$ reperesent the $0.6-10\kev$ flux in the powerlaw and reflection components in  units of $10^{-11}$ $\funit$.
\end{table*}

%%%%%%%%%%%%%%%%%%%%%%%%%%%%%%%%%%%%%%%%%%%%%%%%%%%%%%%%%%%%%%    Nustar Appendix  .......

\appendix

\clearpage

\section{The \nustar{} observation} \label{append:nustar}

\begin{figure*}
  \centering 

\includegraphics[width=6.5cm,angle=-90]{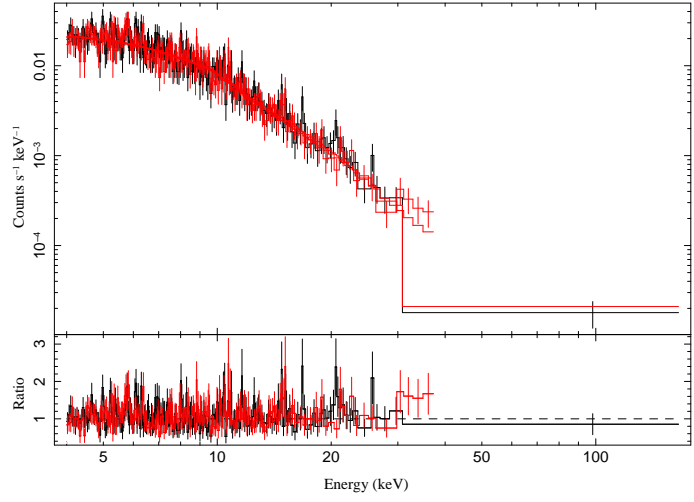} 
	
	\caption{ The \nustar{} FPMA (in red) and FPMB (in black) spectra of the source Mrk~205. The spectra were grouped by a minimum signal to noise ratio of three. {\it Top Panel:} The spectra with the best fit model (a simple powerlaw absorbed by the Galactic column). The best fit statistic is $\rm \chi^2/dof= 376/396 \sim 0.95$. {\it Bottom Panel:} The residuals of the spectra after the fit. See Appendix \ref{append:nustar} for details. The X-axis represents observed frame energy.} \label{fig:nustar1}
\end{figure*}

In this section we discuss the \nustar{} observation (obsid:60160490002) of this source and show why this observation is not suitable for our work, and hence, not used. Mrk~205 was observed by \nustar{} for $\sim 20\ks$ on June 20, 2017. The data were reprocessed using the {\it HEASOFT} command {\it nupipeline} and subsequent spectra for the two instruments FPMA and FPMB were obtained. The source regions were selected using circular regions of $35$ arc-sec radius centred at the centroid of the source coordinates obtained from NED (NASA Extragalactic Database). The background regions were selected using circles of same radius on the same CCD but away from the source. The net counts of the spectra obtained are $\sim 3.5\times 10^3$ for each of the detectors, FPMA and FPMB. The spectra were grouped by a minimum signal to noise ratio of three. Figure \ref{fig:nustar1} shows the spectra in the energy range $4-40\kev$ fitted with a simple powerlaw absorbed by the Galactic column. We found that the spectra is fitted well by this simple model with a $\rm \chi^2/dof= 376/396 \sim 0.95$. The best fit powerlaw slope of $\Gamma=1.84\pm 0.06$, similar to the slope we detected in \suzaku{} and \xmm{} spectra. The \nustar{} spectra is over-modeled even with a simple absorbed powerlaw. We added a Gaussian profile at $6.4\kev$ to model the narrow FeK emission line and the fit improved only by a $\dc=8$ for three parameters of interest. In addition, the normalisation of the Fe line could not be constrained. We therefore, do not use the \nustar{} spectra in our analysis due to lack of statistics in detecting the $\fe$ discrete spectral features and also the broad Compton hump at $\sim 20\kev$. In future we propose to obtain a deeper view of the source with \nustar{}.

$Acknowledgements:$  This research has made use of the NASA/IPAC Extragalactic Database (NED) which is operated by the Jet Propulsion Laboratory, California Institute of Technology, under contract with the National Aeronautics and Space Administration.

\bibliographystyle{mnras}
\bibliography{mybib}

\end{document}